 \documentstyle[pre,aps,floats,twocolumn,epsf]{revtex}

\begin{document}

\def\R{\mbox{$I\!\!R$}}

\def\A{\mbox{\bf A}}
\def\B{\mbox{\bf B}}
\def\D{\mbox{\bf D}}
\def\E{\mbox{\bf E}}
\def\F{\mbox{\bf F}}
\def\G{\mbox{\bf G}}
\def\I{\mbox{\bf I}}
\def\M{\mbox{\bf M}}
\def\P{\mbox{\bf P}}

\def\bfphi{\mbox{\boldmath ${\phi}$}}
\def\bfpsi{\mbox{\boldmath ${\psi}$}}

\def\p{\mbox{\bf p}}
\def\x{\mbox{\bf x}}
\def\y{\mbox{\bf y}}
\def\z{\mbox{\bf z}}

\def\vI{\mbox{\scriptsize {\bf I}}}
\def\vK{\mbox{\scriptsize {\bf K}}}
\def\vzero{\mbox\scriptsize {\bf 0}}

\def\zero{\mbox{\bf 0}}
\def\one{\mbox{\bf 1}}

\draft

\title{Approximating the mapping between systems exhibiting generalized
synchronization}

\author{Reggie Brown}
\address{Department of Physics and Department of Applied Science,
College of William and Mary, Williamsburg, VA 23187-8795}

\date{\today}
\maketitle

\begin{abstract}
We present two methods for approximating the mapping between
two systems exhibiting generalized synchronization.  If the equations of motion 
are known then an analytic approximation to the mapping can be found.
If time series data is used then a numerical approximation can be found.
\end{abstract}

\pacs{05.45.+b}

The subject of synchronization between identical systems (denoted here by IS) has been 
of interest since the time of Huygens.  Over the last decade it has become clear that
even chaotic systems can be synchronized~\cite{chaos}.  One example is 
drive-response synchronization, where
\begin{eqnarray*}
\frac{d \x}{dt} & = & \F(\x) \\
\frac{d \y}{dt} & = & \F(\y) + \E(\x,\y).
\end{eqnarray*}
Here, $\E(\x,\y)$ denotes coupling between the drive system (\x) and the response system
(\y).  If \F\ is deterministic, and if $\E(\x,\x) = \zero$, then the systems are
synchronized if $\y(t_*) = \x(t_*)$.  Because of determinism this condition remains
true for $t > t_*$.  

Recently, papers discussing a more general idea of synchronization have appeared
in the literature.  Drive-response dynamics for this type of synchronization is
given by,
\begin{eqnarray}
\label{d_r}
\frac{d \x}{dt} & = & \F(\x) \\
\frac{d \y}{dt} & = & \G(\y; \x). \nonumber
\end{eqnarray}
where \G\ and \F\ are permitted to be different functions.  In principle, $\x \in
\R^d$, $\y \in \R^r$, and the dynamics takes place in $\R^{d+r}$.  Intuitively, {\em
Generalized Synchronization} (GS) occurs if the response state, \y, is related
to the drive state, \x, by a time independent function, $\y = \bfphi(\x)$.  If GS occurs
then the dynamics takes place on a $d$~dimensional invariant manifold in $\R^{d+r}$.

Much of the work on GS has focused on three areas.  The first area focuses on
defining GS.  Various definitions have been proposed~\cite{avr,kphoy,pjk}.
Reference~\cite{pjk} suggest that subtleties associated with 
unstable periodic orbits imply that more that one definition may be required.  The
second area focuses on mathematical properties of \bfphi.  Rigorous results about
the smoothness of \bfphi, and the relationship between smoothness and Lyapunov exponents
exist~\cite{kphoy,stark}.  Also, numerical methods for determining the properties of 
\bfphi\ exist~\cite{pch}.  Since GS has been observed in experimental 
systems~\cite{rskpp} it is structurally stable.  Mathematical literature 
regarding the existence, stability and smoothness of invariant manifolds is also
relevant~\cite{math}.  The last major area of research has focused on detecting GS 
from time series data~\cite{schiff,rsta}.  The methods are indirect in the sense
that they either do not approximate \bfphi, or the approximations are local (often
resulting in as many approximations as date points). 

This manuscript opens a new research direction.  Instead of seeking properties of
\bfphi, or indirect evidence of its existence, we believe it is better to
go after the function itself.  Therefore, we present methods for analytically
and/or numerically constructing a single smooth function which globally approximates
\bfphi.  If the equations of motion are known then an analytic approximation for 
\bfphi\ can be obtained.  (To our knowledge, this is the only technique for 
analytically approximating \bfphi.)  The numerical method uses time series from
the two systems to calculate a statistic which can be used to infer the existence of 
stable GS.  The numerical method also gives a global approximation for the function,
$\y = \bfphi(\x)$.  (We argue, implicitly, that if \bfphi\ and/or $\bfphi^{-1}$ 
exist but are not well approximated by smooth functions then their usefulness is 
limited since their mathematical properties are probably ``so bad'' they prohibit
most applications of GS.)

An important application for GS comes from control theory.  Typically, control
schemes work better when the complete state of the plant is known.  The
application uses measurements from the plant~(\F) as drive input to an approximate 
model of the plant~(\G).  If GS occurs then the state of the plant can be approximated
from the state of the model via $\x = \bfphi^{-1}(\y)$.  This, and most other
applications, require a stable GS manifold.  

Recently, several criteria have appeared for designing coupling which results in 
a stable IS manifold~\cite{br,gbwm}.  We report here that it is straightforward
to show that a criteria for linearly stable GS is~\cite{new}
\begin{eqnarray*}
\A & \equiv & \left\langle \D_{\bf y} \G[\bfphi(\x); \x] \right\rangle \\
-\Re[\Lambda_1] & > &  \left\langle \| \P^{-1} \left[ \D_{\bf y} \G[\bfphi(\x); \x]
 - \left\langle \D_{\bf y} \G \right\rangle \right] \P \| \right\rangle.
\end{eqnarray*}
Here, $\left\langle \bullet \right\rangle$ denotes a time average over the
driving trajectory, $\Re[\Lambda_1]$ is the eigenvalue of \A\ with the largest
real part, and \P\ is a matrix whose columns are the eigenvectors of \A.  Also,
$\D_{\bf y} \G$ denotes the Jacobian of \G\ with respect to \y.  This criteria
implies that if \bfphi\ and $\bfphi^{-1}$ are known then one can estimate the 
state of the plant~(\x) from the state of the model~(\y) by design coupling which
guarantees stable GS.

The analytical method used to approximate \bfphi\ is based on approximating
center manifolds.  Although the application to GS is new, complete discussions
about approximating center manifolds (with examples) can be found in many
text books~\cite{gh}.  Therefore, our discussion will be brief.  

Assume the drive and response systems are given by Eq.~(\ref{d_r}).  Taking
the total time derivative of $\y = \bfphi(\x)$, and using Eq.~(\ref{d_r}),
implies that
\begin{equation}
\label{PDE}
\G \left[ \bfphi(\x); \x \right] - \left[ \D_{\bf x} \bfphi (\x) \right]
\cdot \F(\x) = \zero 
\end{equation}
on the synchronization manifold.  Here $\D_{\bf x} \bfphi$ is the Jacobian of 
\bfphi.  Equation~(\ref{PDE}) is interpreted as a partial differential equation 
for the unknown function, $\bfphi(\x)$.  The same type of equation arises when
estimating center manifolds~\cite{gh}.  

Typically, Eq.~(\ref{PDE}) can not be solved exactly.  Therefore, approximate
the solution by the series $\bfphi(\x)  = \A + \B \cdot \x + \x \cdot \M \cdot
\x + \ldots$.  Next, insert the series into Eq.~(\ref{PDE}) and rewrite the 
results as a polynomial in powers of \x.  The coefficients of this polynomial 
are functions of the parameters of \F\ and \G\ as well as the elements of \A, 
\B, \M, etc.  Also, this polynomial must hold for all \x\ on the driving 
trajectory.  If this trajectory is not a fixed point then it is reasonable to
assume that the polynomial can hold only if the coefficient of each power of 
\x\ vanishes.  By equating each coefficient to zero we form a set of algebraic
equation involving the parameters of \F, \G, and the elements of \A, \B, \M, etc.
The approximation to $\bfphi(\x)$ is obtained by solving these algebraic equations
for \A, \B, \M, etc in terms of the parameters of \F\ and \G.

Although conceptually straightforward, performing this procedure on anything but the 
simplest examples is very tedious, and soon grows beyond what can be done by hand.  
However, these calculations are not beyond the power of modern symbolic manipulation
software.  Indeed, the results presented below were obtained using MAPLE~\cite{maple}.
It was relatively straightforward to write a MAPLE program which produced these
answers.  Once the program was written the total run time was less than 10 minutes.

The approximation that one obtains for the GS manifold should hold near the 
attractor for the drive dynamics, however, it is not likely to be globally well
defined.  Although the results will not be presented, we have used a similar 
analysis to approximate $\x = \bfphi^{-1} (\y)$ for all of the examples discussed
below.

If the GS manifold is stable then we can numerically approximate \bfphi\ from time 
series data.  The numerical method used to approximate \bfphi\ is similar to one 
used by several authors to make empirical global models from time series data.
Begin by assuming one has two data sets, $\x(n \Delta t) \in \R^d$ and $\y(n 
\Delta t) \in \R^r$, with $n=1, 2, \ldots, N$, which represent simultaneous 
measurements of the drive and response systems at a sampling rate $\Delta t$.
(If necessary, vector representations of the dynamics can be obtained from scaler
time series via embedding techniques~\cite{rmp}.)  A measure for the dynamics
of the drive system can be approximated by~\cite{rmp}
\begin{displaymath}
\rho(\z) = \lim_{N \rightarrow \infty} 
\frac{1}{N} \sum_{n=1}^N \delta [\z - \x(n)]
\end{displaymath}

Since the exact functional form of \bfphi\ is unknown the best one can hope for 
is a series expansion
\begin{equation}
\label{def_phi}
\bfphi(\z) = \lim_{\vK \rightarrow \infty} \sum_{\vI=\vzero}^{\vK} \p^{(\vI)} 
\pi^{(\vI)}(\z).
\end{equation}
Here, the $\p^{(\vI)}$'s are $r$~dimensional expansion coefficients, which must 
be determined, and the $\pi^{(\vI)}(\z)$'s represent some set of basis functions.
Several authors have demonstrated the advantage of using a basis set which is 
orthonormal on $\rho(\z)$, and they show how to construct such a basis from data
using Gramm-Schmidt~\cite{glc,brt}.   The summation index, \I\ is used to
identify the individual basis functions.

Once the basis set has been constructed, each expansion coefficient, $\p^{(\vI)}$,
can be obtained by multiplying both sides of Eq.~(\ref{def_phi}) by $\pi^{(\vI)}(\z)
\: \rho(\z)$ and integrating over all space.  Because of the orthonormality
of the basis set we obtain
\begin{equation}
\label{ps}
\p^{(\vI)} = \lim_{N \rightarrow \infty} \frac{1}{N} \sum_{n=1}^N \y(n) \: 
\pi^{(\vI)}[\x(n)] ,
\end{equation}
where we have used $\y(n) = \bfphi[\x(n)]$ on the GS manifold.  Thus, 
Eqs.~(\ref{def_phi}) and (\ref{ps}) are used to approximate \bfphi\
from time series $\x(n)$ and $\y(n)$.

The last task is to determine the order at which to truncate the series in
Eq.~(\ref{def_phi}) so as to not over fit the data.  This is done by using
the minimum description length (MDL) criteria.  This criteria is similar to 
the maximum likelihood principle associated with least squares fitting of 
data~\cite{brt,jm}.  However, unlike maximum likelihood, MDL is capable of 
determining the optimal order at with to truncate Eq.~(\ref{def_phi}).  The
MDL function we use is given by 
\begin{eqnarray*}
\chi^2_{MDL} & = & \frac{r N}{2} \left[ \ln \left( 2 \pi \hat{\sigma}^2 \right)
+ 1 \right] \\
 & + & N_p \left[ \frac{1}{2} + \ln (\gamma) \right] - \ln(\eta) -
\sum_{\vI = \vzero}^{\vK} \sum_{\beta=1}^r \ln \left( \delta_\beta^{(\vI)} 
\right).
\end{eqnarray*}
(See Ref.~\cite{jm} for a complete derivation of this function.)  Except for
a positive constant (which we neglect), the first term is the usual prediction
error from the maximum likelihood principle.  Indeed, $\hat{\sigma}^2$ is the 
least squares prediction error obtained when predicting the $\y(n)$'s from 
the $\x(n)$'s.

The remaining terms are penalties which increase as more terms in 
Eq.~(\ref{def_phi}) are retained and the model becomes more complex.  $N_p$ 
is the total number of nonzero $\p^{(\vI)}$'s retained in Eq.~(\ref{def_phi}). 
In our implementation, a component of $\p^{(\vI)}$'s is set to zero if its 
statistical significance is not distinguishable from zero~\cite{bit}.  
$\delta_\beta^{(\vI)}$ is the relative accuracy of the $\beta$ component of 
$\p^{(\vI)}$, $\eta$ is the relative accuracy of $\hat{\sigma}^2$, and
$\gamma = 32$.  

To illustrate the analytical and numerical techniques we applied them to 
examples using the Lorenz equations
\begin{eqnarray}
\frac{d x_1}{dt} & = & s(x_2 - x_1) \nonumber \\
\label{lor_drive}
\frac{d x_2}{dt} & = & r x_1 - x_1 x_3 - x_2 \\
\frac{d x_3}{dt} & = & x_1 x_2 - b x_3, \nonumber
\end{eqnarray}
as the drive system.

The coupling between drive and response systems usually involves one of two 
cases.  The first case arises when the physical processes responsible for 
the coupling are known so one has an explicit equation for the coupling.  For
this case one solves Eq.~(\ref{PDE}) as discussed above.  Below we consider
the second case where the response system is given by $\G(\y) + \E [ \bfphi
(\x) - \y]$.  Here, the coupling obeys $\E(\zero) = \zero$ and is used to
insure that the GS manifold is stable.  The problem with this case is that
we can not evaluate $\E[\bfphi(\x) - \y]$ because we do not know, {\em a
priori}, the form of $\bfphi(\x)$.  For the examples discussed below this
problem is overcome by calculating \bfphi\ in two stages.

In the first stage we calculate \bfphi\ using diagonal coupling $\E[\bfpsi
(\x) - \y] \equiv \epsilon [\bfpsi(\x) - \y]$ where $\bfpsi$ is an arbitrarily
function.  The \bfphi\ calculated in this first stage clearly depends on
\bfpsi.  In the second stage we force $\bfpsi(\x) = \bfphi(\x)$.  This second
stage insures that $\E[ \bfpsi(\x) - \y] = 0$ on the GS manifold.

Two trivial tests of the analytic method involved defining $\y = \bfphi(\x) = [x_1 
+ \alpha x_2 + \beta x_2^2, \: x_2, \: x_3]$ for one test, and $\y = \bfphi(\x) = 
[x_1 + \alpha x_3^2, \: \beta x_2, \: x_3 + \gamma x_2^2]$ for the other.   For 
each test we obtain a response system, $\dot{\y} = \G(\y)$, by taking the time
derivative of \y, using Eq.~(\ref{lor_drive}) to resolve the vector field, $\G(\y)$,
and adding diagonal coupling.  For these test, the response systems
are the Lorenz system after a nonlinear change of coordinates, and the analytic 
method easily recovered the GS manifolds, $\y = \bfphi(\x)$.

A final test of the analytic procedure used the following response system
\begin{eqnarray}
\label{lor_response}
\frac{d y_1}{dt} & = & s(1 + \delta) (y_2 - y_1) \nonumber \\
\frac{d y_2}{dt} & = & r(1 + \Delta) y_1 - y_1 y_3 - y_2 \\
\frac{d y_3}{dt} & = & y_1 y_2 - b(1 + \eta) y_3, \nonumber
\end{eqnarray}
with diagonal as discussed above.

For this example we could only approximate \bfphi.  The approximation contained
three arbitrary constant, thus it is not unique.  We selected values for two 
of them so that \bfphi\ has a simple form.  (For the trivial examples discussed
above this choice always lead to the the ``correct'' equation for $\bfphi(\x)$).
The third constant appears trivially in $B_{33}$, is of order $(\Delta, \delta,
\eta)$, and is denoted by $K$ below.  Finally, the approximation is simplified 
by retaining terms that are second order in \x, first order in $(\delta, \Delta,
\eta)$, and in the limit of large coupling strength, $\epsilon$.  (Thus, we
examine a case where the response system is close to the drive system.) 

With these criteria in mind we found that $\bfphi(\x)$ is given by $\A = \zero$,
\begin{displaymath}
\B = \one + \frac{1}{\Gamma} \left[ 
\begin{array}{ccc}
-(r^2 \Delta - s^2 \delta) & -s(r+1-s) \delta & 0 \\
-(r(s-1) \Delta - s^2 \delta) & (r^2 \Delta - s^2 \delta) & 0 \\
0 & 0 & K
\end{array} \right] ,
\end{displaymath}
where \one\ is the identity matrix, and the three tensor \M\ is given by 
$\M^{(1)} = \M^{(2)} = \zero$, and
\begin{displaymath}
\M^{(3)} = \frac{1}{\Gamma} \left[
\begin{array}{ccc}
- \frac{r(s-1) \Delta - s^2 \delta}{b-2s} & 0 & 0 \\
0 & \frac{s (b-2s) (r+1-s) \delta}{b-2} & 0 \\
0 & 0 & 0
\end{array} \right].
\end{displaymath}
In these equations, $\Gamma = (2 r^2 + 3 s^2 - 2s +1)$.  Furthermore, it 
clear that this transformation satisfies $\bfphi = \one$ in the limit 
$\delta, \Delta, \eta \rightarrow 0$.

To test the numerical method we first demonstrate that it can determine the
correct form of \bfphi\ for stable GS from time series data.  To accomplish 
this we used Eq.~(\ref{lor_drive}) (with $s=16$, $b=4$, and $r=46$) as the 
drive system and a response system obtained from $\y = \bfphi(\x) = [x_1 - 
0.01 x_3^2, \: 0.95 x_2, \: x_3 + 0.03 x_2^2]$.  The systems were coupled 
via the $y_2$ equation using $\epsilon [(0.95 x_2 + {\rm noise}) - y_2]$.
The noise was Gaussian white with zero mean and standard deviation, $15 \sigma$.
Here, 15 is approximately the standard deviation of $x_2$, and $\sigma=0$ or
$0.05$.  $\epsilon=10$ was used because, with $y_2$ coupling and a chaotic
driving trajectory, IS is ``stable'' for $\epsilon \geq 4$~\cite{br}.

The numerical procedure was given $N=4000$ simultaneously recorded values 
of $\x$ and $\y$ at a sampling interval of $\Delta t = 0.02$.  The results
(see table~\ref{table}) indicate that the numerical procedure found a good 
approximation to \bfphi, even in the presence of small amounts of noise.
\begin{table}
\begin{tabular}{ccccccc}
 & \multicolumn{3}{c}{$\sigma = 0$} &  \multicolumn{3}{c}{$\sigma = 0.05$}  \\
Factor & $\phi_1$ & $\phi_2$ & $\phi_3$ & $\phi_1$ & $\phi_2$ & $\phi_3$ \\
\tableline
{\rm const}  &  0.00101  &  0.00055  & -0.00194 &  0.0822   &  0.00215 &  0.00475  \\
$x_1$     &  1.00     &  0        &  0       &  1.023    &  0       & -0.0390   \\
$x_2$     &  0        &  0.950    &  0       & -0.0107   &  0.954   &  0.0147   \\
$x_3$     &  0        &  0        &  1.00    & -0.0061   &  0       &  1.003    \\
$x_1 x_1$ &  0        &  0        &  0       &  0.000982 &  0       &  0.000166 \\
$x_1 x_2$ &  0        &  0        &  0       & -0.000304 &  0       & -0.000259 \\
$x_1 x_3$ &  0        &  0        &  0       & -0.000233 &  0       &  0.000381 \\
$x_2^2$   &  0        &  0        &  0.0300  &  0        &  0       &  0.0297   \\
$x_2 x_3$ &  0        &  0        &  0       &  0        &  0       &  0        \\
$x_3^2$   & -0.0100   &  0        &  0       & -0.00994  &  0       &  0
\end{tabular}
\caption{Numerical approximations for the transformation $\phi_1 = x_1 - 0.01
x_2^2$, $\phi_2 = 0.95 x_2$, and $\phi_3 = x_3 + 0.03 x_3^2$.   If the calculate
value of $\phi_j$ was of order $10^{-5}$ or less then it was set to zero.
\label{table}}
\end{table}

To further test the numerical method we used Eqs.~(\ref{lor_drive}) and
(\ref{lor_response}) (the same values for $s$, $b$, and $r$) and a drive signal, 
$\epsilon [(x_2 + {\rm noise}) - y_2]$, coupled to the $y_2$ equation.  These
tests used simultaneously recorded scalar time series of the same length and
sampling interval given above.  Scalars were obtained using the arbitrarily 
chosen projections
\begin{eqnarray*}  
s_d(n) & = & x_1(n) - 2.5 x_2(n) + 0.75 x_3(n) \\
s_r(n) & = & -0.5 y_1(n) + 1.5 y_2(n) - y_3(n).
\end{eqnarray*}
Each scalar time series was independently rescaled to mean zero and standard
deviation one, and an attractor for each time series was reconstructed using 
a time delay embedding~\cite{rmp}.  

\begin{figure}
\vspace{0.0in}
\begin{center}
\leavevmode
\hbox{%
\epsfxsize=3.375in
\epsffile{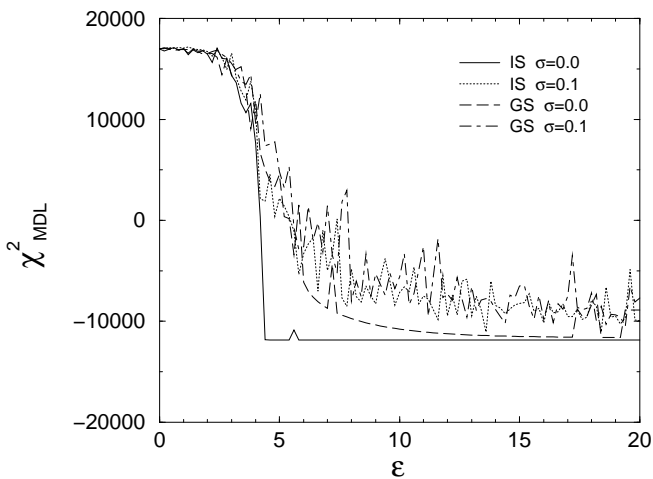}}
\end{center}
\caption{The sudden drop at $\epsilon \simeq 4$ indicates the onset of stable
synchronization and stable generalized synchronization.
\label{figure}}
\end{figure}

The results of our attempts to approximate \bfphi\ for $\delta = \Delta = \eta
= 0$ (IS) and $\delta = 0.02$, $\Delta = 0.04$, $\eta =-0.03$ (GS) are shown in
Fig.~\ref{figure}.  The figure shows that $\chi^2_{MDL}$ experiences a sharp 
drop at $\epsilon \simeq 4$ when the drive/response systems are identical and 
a less sharp drop for GS.  The drop implies that the numerical procedure has 
found a relatively accurate approximation for $\y = \bfphi(\x)$, so the GS 
manifold is stable.  Also, the figures shows that the procedure deteriorates 
gracefully in the presence of noise.

In conclusion, we have presented an analytical and a numerical method for 
approximating the mapping that defines the invariant manifold associated with
generalized synchronization.  The author would like to thank Drs. N. F. Rulkov,
L. M. Pecora, B. R. Hunt, and J. Stark for valuable discussions and comments 
that lead to this work.


\begin{references}
\bibitem{new} This result is easily obtained by applying methods discussed
in Ref.~\protect{\cite{br}}.
\end{references}
\end{document}